\documentstyle[11pt,aaspp4]{article}

\begin{document}

\title{Optical Spectroscopy of Diffuse Ionized Gas in M31}
\author{B. Greenawalt and R.A.M. Walterbos\altaffilmark{1}}
\affil{Astronomy Deparment, Box 30001, Department 4500, New Mexico 
State University, Las Cruces, NM 88003}
\authoremail{bgreenaw@nmsu.edu}

\and

\author{R. Braun\altaffilmark{1}}
\affil{Netherlands Foundation for Research in Astronomy,
Radiosterrenwacht, P.O. Box 2, 7990 AA Dwingeloo, The Netherlands}

\altaffiltext{1}{Visiting Astronomer, Kitt Peak National Observatory,
National Optical Astronomy Observatories, which is operated by the
Association of Universities for Research in Astronomy (AURA) under
cooperative aggreement with the National Science Foundation.}

\begin{abstract}

We have obtained sensitive long-slit spectra of Diffuse Ionized Gas
(DIG) in the Andromeda Galaxy, M31, covering the wavelength range of
3550-6850$\rm\AA$.  By co-adding extracted DIG spectra, we reached a
1$\sigma$ uncertainty of $\rm 9.3\times10^{-19}~ergs~s^{-1}~cm^{-2}~
arcsec^{-2}$ corresponding to .46 pc cm$^{-6}$ in Emission Measure. We
present average spectra of DIG at four brightness levels with Emission
Measures ranging from 9 to 59 pc cm$^{-6}$.  We present the first
measurements of [OII]$\lambda3727$ and [OIII]$\lambda5007$ of the
truly diffuse ionized medium in the disk of an external spiral galaxy.
We find that $\rm I_{[OII]}/I_{H\alpha}=.9-1.4$. The [OIII] line is
weak ($\rm I_{[OIII]}/I_{H\beta}$ = .5), but it is stronger than found
for the Galactic DIG.  Measurements of [NII]$\lambda6583$ and
[SII]($\lambda6717+\lambda6731$) are also presented. The [SII] lines
are clearly stronger than typical HII regions ($\rm
I_{[SII]}/I_{H\alpha}$ = .5 compared to .2), confirming various
imaging studies of spiral galaxies.  Overall, the line ratios are in
agreement with predictions of photoionization models for diffuse gas
exposed to a dilute stellar radiation field, but the line ratios of
the DIG in M31 are somewhat different than observed for Galactic DIG.
The differences indicate a less diluted radiation field in the DIG of
M31's spiral arms compared to DIG in the Solar Neighborhood of the
Milky Way.  Turbulent mixing layers can contribute at most 20\% of the
ionization budget of the DIG, with lower percentages producing better
fits to the observed line ratios.  We have also detected HeI $\lambda
5876$ emission from the brightest DIG in M31. The HeI line appears to
be stronger than in the Galactic DIG, possibly indicating that most of
the Helium in the bright DIG in M31 is fully ionized. However, this
result is somewhat tentative since bright night sky lines hamper an
accurate measurement of the HeI line strength.

\end{abstract}


\section{Introduction}

For years, it has been known that extensive quantities of ionized
hydrogen, not restricted to the immediate vicinity of high mass stars,
exists in the Milky Way and several external galaxies.  In the Galaxy,
this gas is referred to as the Reynolds Layer (Reynolds
\markcite{rey91}1991), while a more general description is Diffuse
Ionized Gas (DIG) or the Warm Ionized Medium (WIM). DIG is seen as a
thick layer above the disk of some edge on galaxies (eg. NGC 891: Rand
\markcite{rand96}1996, Rand et al. \markcite{rkh90}1990, Dettmar
\markcite{det90}1990, Dettmar \& Schulz \markcite{ds92}1992, Keppel et
al. \markcite{kdgr91}1991; NGC 4631: Dettmar \markcite{det92}1992,
Rand et al. \markcite{rkh92}1992, Golla et al. \markcite{gdd96}1996;
and NGC 55: Ferguson et al. \markcite{fwg96a}1996a), as bright
filaments in some irregular galaxies (Hunter \& Gallagher
\markcite{hg96}1996, \markcite{hg92}1992, \& \markcite{hg90}1990,
Hunter \markcite{hunt96}1996, \markcite{hunt94a}1994a, Hunter et
al. \markcite{hhg93}1993), and as both diffuse and filamentary regions
in several nearby ``face-on'' spiral galaxies (M31: Walterbos \& Braun
\markcite{wb94}1994, Hunter \markcite{hunt94b}1994b; M33: Hester \&
Kulkarni \markcite{hk90}1990, Greenawalt \& Walterbos
\markcite{gw97a}1997a, Hunter \markcite{hunt94b}1994b; M51 \& M81:
Greenawalt \& Walterbos \markcite{gw97b}1997b; Sculptor Group
galaxies: Hoopes et al. \markcite{hwg96}1996, Ferguson et
al. \markcite{fwgh96b}1996b). In all cases, most of the DIG is found
to be concentrated in regions of star formation, although isolated
patches and filaments are found as well.

While observations of this component of the ISM have occurred with
increasing frequency over the last 5 years, the ionization and overall
3-D morphology are still poorly understood.  Various models have been
suggested to account for the ionization: Lyman continuum photons
emitted by OB stars (Mathis \markcite{math86}1986, Domg$\rm\ddot
o$rgen \& Mathis \markcite{dm94}1994, Dove \& Shull
\markcite{ds94}1994, Miller \& Cox \markcite{mc93}1993), turbulent
mixing of hot and cool gases (Slavin et al. \markcite{ssb93}1993),
shocks from supernovae (see Reynolds \markcite{rey84}1984), and
decaying dark matter (Sciama \markcite{sci90}1990,
\markcite{sci93}1993, \& \markcite{sci95}1995).  Each of these models
has its difficulties, with some problems being more severe than
others.  Photoionization models need large mean free paths for
ionizing radiation, requiring the ISM to be perhaps more porous than
previously thought.  However, these models are most attractive from an
energetics standpoint, although still requiring up to 40\% of the
ionizing radiation from OB stars.  Conversely, shocks from supernovae
and mixing layer models are unable to ionize all of the DIG, even if
all the energy of supernovae were available.  Also, emission line
ratios of the emitting gas constrain the shock speeds to an
uncomfortably narrow range ($\sim 80\pm10$ km/s) (Raymond
\markcite{ray79}1979, Shull \& McKee \markcite{sm79}1979).  Decaying
dark matter models require other sources of ionizing radiation to
produce [NII]($\lambda6548+\lambda6583$), HeI($\lambda5755$), and
[OIII]($\lambda4959+\lambda5007$) (Sciama \markcite{sci95}1995), and
should be considered speculative.

Observations of DIG in external galaxies have largely been
accomplished with imaging of H$\alpha$ and
[SII]($\lambda6717+\lambda6731$), and recently, [OII]$\lambda$3727,
the blend of [OII]($\lambda3726+\lambda3729$), through narrow band
(25-75$\rm\AA$) filters (eg. Walterbos \& Braun \markcite{wb92}1992,
\markcite{wb94}1994; Ferguson et
al. \markcite{fwg96a}\markcite{fwgh96b}1996a,b; Rand et
al. \markcite{rkh90}1990, \markcite{rkh92}1992).  Spectroscopic
studies have been mostly confined to thick ionized layers above
edge-on systems (Rand \markcite{rand96}1996, Golla et
al. \markcite{gdd96}1996) and filaments in irregulars (e.g. Hunter and
Gallagher \markcite{hg96}1996).  The imaging studies have shown that a
surprisingly consistent fraction (30-50\%) of the total H$\alpha$
emission from spirals arises in the DIG.  Some recent studies have
suggested that the [SII]/H$\alpha$ intensity ratio increases with
decreasing surface brightness. To better test ionization models,
observations of other lines are important. While the observations of
edge-on systems address the ionization and excitation state of the gas
{\it above} galactic disks, they do not provide information on the
line ratios {\it in} the galactic disks.

Here we present the initial results from a spectroscopic study of the
emission from faint DIG in the non-edge-on spiral M31.  The spatial
resolution provided by M31's proximity allows clear separation of DIG
regions from HII regions.  This lack of ``confusion'' aids in the
observation of faint DIG.  Previous sensitive H$\alpha$ imaging
(Walterbos \& Braun \markcite{wb92}1992, \markcite{wb94}1994) over
most of the NE half of the galaxy provided valuable complimentary
information and aided the acquisition and interpretation of these
spectroscopic data.

In \S 2 we review the observational procedures and data reduction
techniques.  The observed forbidden line ratios are presented in \S 3,
along with the detection of HeI and temperature estimates.  In \S 4 we
compare the predictions of ionization models to our observations.  In
\S 5 we discuss the observed characteristics of DIG in M31 to that in
other galaxies.  Finally, \S 6 contains a summary of our results.

\section{Observations and Data Reduction}

The spectra were obtained using the RC spectrograph on the Mayall 4-m
telescope at Kitt Peak National Observatory on Nov. 3-5, 1991.  The
detector was a Tek CCD chip with 1024x1024 pixels, each 0.69
arcseconds (2.3 pc for M31) on a side. A slit of about 6 arcminutes
length and 2 arcseconds width was used.  The slit was rotated to
provide the best compromise between the ideal parallactic angle and
the desire to include many interesting objects on the slit, eg. HII
regions, SNRs, and regions of diffuse ionized gas.  These objects were
identified on H$\alpha$ and [SII] images obtained by Walterbos and
Braun \markcite{wb92}(1992).  Sixteen different slit positions were
observed in the NE half of M31.  They are mainly concentrated in the
spiral arms, over a range in galactic radii of 5 to 15 kpc, with a
concentration of slit locations in the 10 kpc star formation
``annulus''.  The locations of the slits on the sky will be presented
elsewhere, when we discuss the results for the individual spectra
(Greenawalt et al. 1997); here we are concerned only with the average
DIG spectra at various brightness levels. We used the KPC-10A grating,
with 316 lines/mm which is blazed to provided maximum sensitivity at
4000$\rm\AA$ in first order.  The final dispersion with this grating
was 2.77$\rm\AA$/pixel.  For each slit position the grating was
toggled between two tilt angles, centering the spectra on 4950$\rm\AA$
and 5500$\rm\AA$, to provide a blue and a red spectrum.  This strategy
provided wavelength coverage from 3550-6850$\rm\AA$ with considerable
overlap.  At each tilt angle, two 15 minute object spectra and one
wavelength calibration spectrum were taken.

Standard IRAF tasks were used to reduce the data. The CCD DC offset
was removed by fitting the overscan regions on each spectrum.  Any
residual bias structure was removed using a combination of bias images
taken each night.  Pixel-to-pixel variations were corrected using
spectra of a quartz lamp illuminating the dome.  Cosmic rays were
edited out manually to insure that no spectral features were altered.

Each pair of spectra for a particular slit position and central
wavelength were averaged together.  The wavelength solution for the
matching spectrum of a HeNeAr arc lamp was determined, then applied to
the object spectrum.  The spectra were flux calibrated using a
standard star spectrum obtained during the night.  One standard star
observation from each night was used to calibrate the entire nights
data.  The sky conditions were clear, but we did not rigorously verify
if conditions were photometric. Nevertheless, the emission measures we
derive on the average spectra (see below) agree well with those
derived from the corresponding image sections on the H$\alpha$ images
obtained by Walterbos \& Braun \markcite{WB94} (1994).

The detector focus was slightly poorer at the edge of the field
compared to the center of the field.  A spectral line was therefore
broader and the peak brightness was lower at the edge of the field.
Because of this structure in the night sky lines, fitting a constant
or a linear fit to a night sky line resulted in unacceptably large
residuals when attempting a standard sky subtraction. Tests showed
that a 7th order polynomial fit provided the best subtraction of the 4
brightest night sky lines, $\rm[OI]\lambda5577,
NaI(blend)\lambda\lambda5890+5896, [OI]\lambda6300, and
[OI]\lambda6364$. Since these lines have no underlying M31 nebular
emission, such a high order fit was well constrained.  Rmoval of the
other night sky lines is more complicated, however.

The strengths of individual night sky lines vary in differing amounts
from one spectrum to the next. Thus even if we had observed separate
blank-sky spectra, we would not have obtained good removal of all
night sky lines in individual M31 spectra. The standard technique, of
fitting on background regions in each individual spectrum could only
be used in a limited way, since the emission from DIG in M31 extended
over most of the slit, severely limiting the choice of ``background''
regions.  In addition, with only a few background regions per spectrum
a high order fit to the sky features was not possible.  Therefore, we
were forced to develop an alternative background subtraction method.
This method took advantage of the fact that the structure of a night
sky line did not vary as the strength of the line changed.

In what follows it may be useful to remember that the dispersion
direction is sometimes referred to as ``columns'' or ``vertical
direction'', while the spatial direction along the slit is referred to
as ``rows''. All spectra were shifted to a common wavelength grid.
First, the four brightest night sky lines were removed from the
individual spectra, using the following method. The smoothly varying
continuum in each spectrum was removed by subtracting a median
filtered version of the original spectrum.  Continuum sources, such as
foreground stars \& M31 emission, were removed as well in this step,
leaving only the individual fainter night sky lines and the nebular
emission features in the spectra. The smoothly varying continuum of
the sky brightness was determined by applying a rectangular median
filter box of 1$\times$151 pixels, with the long dimension aligned
with the dispersion axis. On the spectrum with the continuum
subtracted, a 7th order 1-D polynomial was fit to each of the rows
containing emission from the four bright sky lines.  These fits were
then removed from the original spectrum.  The method was iterated
until the sky line subtraction produced no noticeable depressions in
the continuum level at the wavelength of the lines. Removing the
brightest lines first was required to avoid the creation of large dips
near the locations of these lines when subtracting the median filtered
image.  All subsequent steps were performed on the spectra from which
the four bright night sky lines had been removed.

Next, we determined templates of the remaining night sky lines, which
could then be used to remove these lines from the spectra. For this
purpose, we selected the two spectra with the least amount of nebular
emission from M31, hence with the most ``background'' regions on which
the night sky lines could be fit. We again first removed the smoothly
varying continuum, as described above. On each spectrum, we then
selected background regions, to which 3rd order 1-D polynomials were
fit, row by row. Higher order fits were not possible given the extent
along the slit of the background regions available.  The polynomial
fits to each spectrum were written out as a template spectrum of night
sky lines. The resulting two templates were averaged together to
produce a final night sky line template. 

The final step consisted of removing the night sky lines from all
spectra using this template. To do this, we again determined the
smoothly varying continuum for each spectrum using the median filter
as described above, and subtracted this from the original spectrum. We
then identified background regions on all spectra. For each row, we
determined the mean intensity in the background regions, and the
corresponding mean intensity in the night sky template for the same
regions. This provided the scaling factor required to create an
individual night sky template for each spectrum; this was then
subtracted from the original spectrum, with the continuum still in it,
to produce the final M31 spectra. Even in cases where the emission
filled most of the slit, good sky subtraction was obtained this
way. By subtracting the templates from the original spectra, no
potential artifacts introduced by the median filtering can affect the
data. With the night sky lines removed, only a smoothly varying
background component, due to continuum emission from M31 and the night
sky, was left to remove.  A linear fit to the background regions along
each row proved sufficient to remove this final background component.

\section{Results}

\subsection{Representative DIG spectra}

After the background was subtracted, we identified the various
emission-line objects covered by each slit location, using the narrow-band
images of the fields \markcite{wb92}(Walterbos and Braun 1992).  Based
on morphology and brightness level, we classified 54 apertures as
containing DIG.  These ranged from roughly 10 to 60 pixels in extent,
with a mean value of 23.5 pixels (54 pcs).  Apertures containing HII
regions, Supernova Remnants, candidate Luminous Blue Variables,
Wolf-Rayet stars and other objects were also extracted.  These will be
discussed elsewhere (Walterbos et al. \markcite{wk96}1996, Galarza et
al. \markcite{gwg97}1997).

All extracted spectra were shifted so the emission lines were at their
respective rest wavelengths.  This step removed the systemic velocity
of M31 and any rotational velocity differences between the DIG
locations.  Using gaussian fits to line profiles in the task SPLOT,
the H$\alpha$ flux was measured for each DIG aperture and then
converted to average intensity using the number of summed pixels in
each aperture.  The Emission Measures (EM) derived from the H$\alpha$
surface brightness, assuming T$\rm_e = 10^4$K, of the DIG within the
54 apertures ranged from 12-71 pc cm$^{-6}$. After ordering the
apertures by decreasing surface brightness, they were separated into 3
groups of 18 apertures each.  The 1D spectra within each group were
then combined to produce representative spectra of ``Bright'',
``Moderate'', and ``Faint'' DIG, with EMs of 58.4, 33.8, and 17.7 pc
cm$^{-6}$, respectively.  In addition, twelve DIG regions located well
away from any bright HII regions or filaments were combined to produce
a representative spectrum of DIG ``Far'' from HII regions. This DIG
had an average EM = 22.6 pc cm$^{-6}$.  To compliment these spectra,
we extracted 26 apertures of regions with very faint emission which
showed only marginal detection of H$\alpha$ in individual cases.
These were shifted using adjacent apertures of brighter emission as
reference, then combined to produce a representative spectrum of the
``Very Faint'' DIG with emission measure of 9.1 pc cm$^{-6}$.  The
1$\sigma$ noise in all final co-added spectra was roughly $\rm
9.3\times10^{-19}~ergs~s^{-1}~cm^{-2}~arcsec^{-2}$, corresponding to
an EM of .46 pc cm$^{-6}$. This noise level is low compared to the
faintest DIG we extracted from the spectra, mainly because the noise
in individual spectra is much higher, so identifying regions at these
faint levels is not trivial. In addition, since emission may fill most
of the slits completely, extracting fainter spectra introduces a
systematic uncertainty in the derived line ratios due to possible
subtraction of a low-level (estimated by Walterbos \& Braun
\markcite{WB94} (1994) to be no higher than EM = 3 pc cm$^{-6}$)
genuine DIG component. The low noise in the co-added spectra implies
that we have excellent S/N even for the ``Very Faint'' DIG.

Table 1 contains the line ratios for each of these combined spectra.
The quoted ``uncertainties'' in the surface brightness of the combined
spectra correspond to the 1$\sigma$ rms spread in the distribution of
surface brightnesses of the individual spectra.  In the case of the
``Very Faint'' DIG, this number is an estimate, since surface
brightnesses for individual spectra were not measured.  Uncertainties
quoted for line ratios reflect 1$\sigma$ errors determined from the
combined spectra.  We will report on results for the individual
spectra in a different paper (Greenawalt et
al. \markcite{gwb97}1997). It is important to point out that the
intrinsic spread in the line ratios from one region to another is much
larger than the quoted uncertainties here, which are derived from the
co-added spectra. Thus there is significant variation in line ratios
across the DIG but globally the line ratios vary little.

The measured emission line intensities from each combined spectrum
were corrected for reddening using the Balmer decrement and a standard
Galactic extinction curve (Savage \& Mathis \markcite{sm79}1979).  The
ratio $\rm I_{H\alpha}/I_{H\beta}$ was 4.4 for the ``Very Faint'' DIG
spectrum and roughly 3.6 for others, compared to an unreddened value
of 2.86 for 10$^4$ K gas.  Because the absorbing material may well be
mixed with the DIG, we calculated corrected line ratios for both a
foreground screen model and a model consisting of a Galactic
foreground dust screen and a homogeneous mixture of dust and DIG in
M31.  For a foreground dust screen, the extinction implied by the
observed line ratios is A$\rm_V \simeq$ .7 (A$\rm_V \simeq$ 1.2 for
the ``Very Faint'' DIG), roughly twice the foreground Galactic
extinction, for which E(B-V) = .1 (e.g., Walterbos \& Schwering
\markcite{ws1987}1987).  The model, with a component of dust and DIG
mixed in M31, suggests the dust column from this component alone would
correspond to an A$\rm_V \simeq$ .6-1.1 if it were a foreground screen
(A$\rm_V = 3.1$ for the ``Very Faint'' spectrum), in addition to the
foreground Galactic extinction, A$\rm_V = .3$.  Because of the dust
distribution in this model, the corrected line ratios are actually
slightly smaller compared to a model of only foreground extinction, in
spite of the increased estimation of the total extinction.  In the
most extreme case of $\rm I_{[OII]}/I_{H\alpha}$, the line ratios are
reduced by less than 5\% (less than 10\% for the ``Very Faint''),
roughly consistent with our 1 $\sigma$ error estimation.  Because the
exact distribution of dust has little effect on the corrected line
ratio in this case, we quote corrected line ratios assuming the
extinction arises within a foreground dust layer.  The reddening
corrected ratios of other Balmer line ratios, in particular,
$\rm{H\gamma}\over{H\beta}$ and $\rm{H\delta}\over{H\beta}$ for all
DIG spectra are consistent with the expected values for unreddened
10$^4$ K ionized gas.

\subsection{Forbidden line ratios}

Portions of the observed spectra around H$\alpha$ are shown in Fig. 1.
The figure gives a good indication of our signal-to-noise and
wavelength resolution. Night sky lines just blueward of H$\alpha$
caused some problems with the accurate determination of both the
strength of the [NII]$\lambda6548$ line and the baseline level; all
[NII] measurements therefore refer to the $\lambda 6584$ line.  We
find a constant $\rm I_{[NII]\lambda6583}/I_{H\alpha}$ ratio of .35
for all four brightness levels of DIG, but a slightly higher value of
.40 for DIG ``Far'' from HII regions.  This is similar to that
measured for HII regions and for the Reynolds layer in our Galaxy
(Reynolds \markcite{rey90}1990).

The strength of the [SII]$\lambda6717+\lambda6731$ lines relative to
the H$\alpha$ line has been recognized for some years to be different
in DIG versus HII regions.  Many imaging programs have used the value
of $\rm I_{[SII]}/I_{H\alpha}$ along with the brightness of H$\alpha$
to distinguish DIG in nearby galaxies (e.g. Walterbos and Braun
\markcite{wb94}1994, Hoopes et al. \markcite{hwg96}1996, Ferguson et
al. \markcite{fwgh96b}1996b).  Our spectra show values of $\rm
I_{[SII]}/I_{H\alpha}$ in the range of .40-.50, which is significantly
larger than the average value of .25 observed in HII regions
(Walterbos and Braun \markcite{wb94}1994, also Fig. 4).  The ratio
obtained for the ``Far'' DIG is .60, considerably higher.  There
appears to be a trend of increasing ratio with decreasing surface
brightness, both in the combined spectra and in the individual spectra
before combination (Greenawalt et al. \markcite{gwb97}1997).

Figure 2 shows a large portion of the blue spectra.  Wavelengths
shortward of 4200$\rm\AA$ were observed only in the blue spectra, and
therefore have lower signal-to-noise.  The decline of detector
sensitivity and reduced atmospheric transmission in this wavelength
region also have a negative effect on the signal-to-noise.  In spite
of these facts, we clearly detect the [OII] doublet at $\lambda3727$
in all DIG spectra.  The reddening corrected $\rm
I_{[OII]}/I_{H\beta}$ ratio is 2.6 for the ``Bright'' and ``Moderate''
spectra, 3.4 for the ``Faint'' spectrum, and 4.1 for the ``Very
Faint'' spectrum.  This translates to $\rm I_{[OII]}/I_{H\alpha}$ =
.9, 1.2, and 1.4, respectively.  The ratio obtained for the ``Far''
DIG is basically the same as for the ``Faint'' DIG.  The
[OIII]$\lambda4959$ line is weak and not seen in the spectra of the
faintest DIG, but the stronger of the doublet, [OIII]$\lambda5007$, is
accurately measured in all spectra.  The $\rm
I_{[OIII]\lambda5007}/I_{H\beta}$ ratio for DIG in M31 ranges from .35
to .60.  Note that the [OI]$\lambda$6300 line, if present, could not
have been detected in M31 because of the bright [OI]$\lambda$6300
night sky line.  The hydrogen lines H$\beta$, H$\gamma$, and H$\delta$
were also detected in the three brightest DIG spectra.

\subsection{Search for HeI recombination line}

The observations of strong [SII]($\lambda6717+\lambda6730$) and
[NII]($\lambda6548+\lambda6583$) emission coupled with weak
[OI]$\lambda6300$ and [OIII]$\lambda5007$ emission from the Reynolds
layer within our Galaxy and the thick ionized layer surrounding NGC
891, imply that low ionization species dominate.  While some
ionization models are able to reproduce the observed line ratios,
the hardness of the incident ionizing spectrum is a key to
distinguishing between models.  The HeI$\lambda5876$ recombination
line is sensitive to the spectral type of the ionizing star, since
ionization of He requires stars of spectral type earlier than about O8
(Torres-Peimbert et al. \markcite{tlp74}1974).  This emission line is
the strongest helium line in the visible regime, but is still
extremely weak compared to H$\alpha$.  

Fig. 3 shows the region around HeI$\lambda5876$ in the spectrum of the
``Bright'' and ``Moderate'' DIG.  The figure shows the red spectra
prior to combination with the blue spectra.  Because this region is
close to the edge of the chip in the blue spectra, background
subtraction was more difficult resulting in larger residuals for the
night sky lines around HeI.  The red spectrum of the ``Bright'' DIG
shows a detection of HeI$\lambda5876$.  There is a feature in the blue
spectrum consistent with the detection of HeI at the same strength as
in the red spectrum, even though the night sky lines residuals are
greater.  The bright NaI$\lambda5890+\lambda5896$ night sky lines just
redward of HeI did not subtract perfectly, producing the obvious dip
in baselevel.  Although there are shifts of the emission features in
the original spectra relative to the night sky lines amounting to
about 5$\rm\AA$ because of M31's galactic rotational velocity
differences for various slit positions, the HeI line is always
separated from this bright night sky line.  However, a weaker OH
molecular blend is blueward of HeI and sometimes blended with the
line.  This blend produces a broad pedestal in the ``Moderate'' DIG
spectrum where we do not detect the HeI line, also shown in fig. 3.
Because there is the possibility that this fainter night sky line
could contribute to our estimate of the strength of the HeI line via a
slight hump on the blue side of the HeI line, we consider our
determination of the HeI line flux to be uncertain.  To avoid a
possible contribution from this hump we estimated the strength of the
HeI line from the {\it peak} line intensity for HeI compared to
H$\alpha$, as opposed to comparing total line fluxes.  The measured
ratio of $\rm I_{HeI}/I_{H\alpha}$ = .045$\pm$.015.  The upper limits
from the other spectra are still consistent with this level of
ionization, but, of course, also with lower levels of ionization of
He.  The sky subtraction produced sky line residuals in this region of
the red ``Faint'' and ``Very Faint'' spectra consistent with other
regions, although not shown in fig. 3.

Our estimated ratio for $\rm I_{HeI}/I_{H\alpha}$ is consistent with
the value found for Galactic HII regions, .045 (Reynolds \& Tufte
\markcite{rt95}1995). This implies that stars of spectral type earlier
than O7 are required to contribute to the ionization of the DIG.
Reynolds \& Tufte unsuccessfully searched for helium emission from two
directions through the Reynolds layer.  Their 2$\sigma$ upper limit of
$\rm I_{HeI}/I_{H\alpha}$ = .012$\pm$.006 implies that He is primarily
neutral and the radiation field is softer than previously thought.
The two directions observed have surface brightness levels of about 30
pc cm$^{-6}$, the same as our ``Moderate'' spectra.  Since, we are
only able to place upper limits on $\rm I_{HeI}/I_{H\alpha}$ for the
``Moderate'', ``Faint'' and ``Very Faint'' DIG, the question remains
open as to the hardness of the radiation field responsible for the
fainter DIG.  However, it is appears that the radiation field present
in the ``Bright'' DIG of M31 is harder than in Galactic DIG.  This may
result from the fact that our observations focus on DIG regions within
the spiral arms of M31, while the Galactic observations sampled DIG
farther from spiral arm structures.

Recent observations of ionized gas out of the plane of NGC 891 have
found $\rm I_{HeI}/I_{H\alpha}$ = .033 (Rand \markcite{rand96}1996).
This ratio suggests that helium is about 70\% ionized and the
equivalent stellar temperature of the radiation field is 37,500K,
corresponding to spectral type O7.  The consistency of these
measurements with our estimation for the ``Bright'' DIG in M31 is
somewhat surprising, especially given the discrepancy with Galactic
DIG.  Given that NGC 891 and the Milky Way are considered ``twins'',
with large ionized layers above the plane of each, one might expect the
HeI line strength to be similar in the extraplanar DIG for these
galaxies and possibly different for the DIG located within the plane of
M31.

Also located in this spectral region is the [NII]$\lambda5755$ line.
This line is expected to be extremely weak, but important for
temperature determination when compared to [NII]$\lambda 6548+\lambda
6583$.  Equation 5.5 of Osterbrock \markcite{ost89}(1989), in the
low-density limit (which is confirmed by [SII] lines), reads:
$$ \rm {{[NII]\lambda 6548 + [NII]\lambda 6583}\over {[NII]\lambda
5755}} = 6.91 exp \Biggl( { {2.5 \times 10^4}\over{T}} \Biggr) . $$
While we have no clear detection of the [NII]$\lambda 5755$ line even
in the ``Bright'' DIG spectrum, we can estimate an upper limit to the
temperature using the measured noise at the line position.  The
measured noise gave a 3$\sigma$ upper limit on the line flux of $\rm
9.90\times10^{-16}~erg~s^{-1}~cm^{-2} $.  This noise limit does not
place a strong constraint on the temperature of the DIG, but does
suggest it is most likely below 17,000K.

\section{Ionization Models \& Discussion}

By comparing the predictions of several ionization models and
observations of DIG in the Galaxy to the observations of DIG in M31,
we will attempt to address its ionization source.  The pertinent line
ratios from Galactic DIG and models discussed below are included in
Table 1 for comparison with the observed line ratios discussed above.
Figures 4 and 5 are excitation diagrams which are useful in
determining the model best able to fit the observed line ratios for
DIG.  Line ratios for several HII regions from our data set (Galarza
et al. \markcite{gwg97}1997) have been included to illustrate the
striking differences from DIG regions.  In particular, the
[SII]($\lambda6717+\lambda6731$)/H$\alpha$ ratio is at least a factor
of 2 greater in the DIG compared to the compact HII regions, and the
[OII]$\lambda3727$/H$\alpha$ ratio on average is higher in the DIG
than the compact HII regions.

\subsection{Photoionization}

At least for the solar neighborhood, Miller \& Cox
\markcite{mc93}(1993) have shown that the observed O star distribution
can supply the required ionizing radiation to produce the observed
Galactic DIG.  Mathis \markcite{math86}(1986) and Domg$\rm\ddot o$rgen
\& Mathis \markcite{dm94}(1994) (hereafter DM) produced
photoionization models based on the idea that diffuse stellar
radiation is able to travel through a low-density ISM to ionize H far
from high mass stars.  The governing model parameters are X$_{edge}$,
the fraction of neutral hydrogen at the model edge, and q, a measure
of the mean ionizing photon density within the nebula to the mean
electron density.  Their ``composite'' models are a mixture of 20\%
X$_{edge}$=.95 gas (ie. edges of HI clouds), and 80\% X$_{edge}$=.1
gas, which produces much of the H$\alpha$ emission.  In this
``composite'' model, most of the emission is attributed to a low
X$_{edge}$ model because only these models produce the low $\rm
I_{[OI]}/I_{H\alpha}$ ratio observed in the Galactic DIG.  Composite
models were only published for log(q) in the range -3 to -4, which
appears appropriate for the Galactic DIG.  These composite models fit
the observed $\rm I_{[NII]}/I_{H\alpha}$ and $\rm
I_{[OII]}/I_{H\beta}$ ratio, although the observed values of [OII]
seem to have a larger scatter than model predictions (see fig 5).  As
is obvious from fig. 4, these composite models seem to slightly
overpredict [SII] while underpredicting [OIII].  The composite models
were created to reproduce observations of the Galactic DIG, which has
a low $\rm I_{[OIII]}/I_{H\beta}$ ratio.  Mathis
\markcite{math96}(1996) has suggested that increasing [OIII] in model
predictions would be easy, since it was difficult to bring down the
ratio to match Galactic DIG observations.  Domg$\rm\ddot o$rgen \&
Mathis also published results for X$_{edge}$=.10 models alone, with a
much larger range of q values.  The $\rm I_{[OII]}/I_{H\alpha}$ ratio
was only published for log(q) = -3, -3.7, and -4 models, and showed a
variation of less than 1\% for these models.  We therefore estimated
the strength for other models to be the same.  A model with q=.005
fits our observed line ratios the best, but values from .002 to .01
are consistent with the data.  The larger q values imply a larger
ratio of ionizing photons to gas density.  Because the observed
surface brightness of M31 DIG is in the same range as Galactic DIG
(5-35 pc cm$^{-6}$), we are not detecting systematically brighter DIG
in M31.  The difference in line ratios suggests that the conditions in
M31 are somewhat different than in the Milky Way.  The DIG regions we
observed are primarily in spiral arms, while the Galactic DIG regions
are not.  The less diluted radiation field required to fit our
observation may reflect the fact that there is a more prominent
stellar radiation field within spiral arms.  This is supported by the
HeI detection in M31 compared to the upper limit found for Galactic
DIG.  We expect that the larger q value is the important factor in
finding a model that fits our observations, not the specifics of
``composite'' or single X$_{edge}$ model.  Because we have no
meaningful limits on [OI], a low X$_{edge}$ model is not necessarily
required.  Results for a ``composite'' model for these larger q values
would be valuable.

As can be seen in fig. 4, the DM models predict an increase in the
$\rm I_{[SII]}/I_{H\alpha}$ ratio with decreasing q parameter.  If the
q parameter appropriate for a given DIG region is related to the
surface brightness of the region, then the models predict an increase
in $\rm I_{[SII]}/I_{H\alpha}$ with decreasing surface brightness.
Just such a trend is observed in our representative spectra, and the
individual spectra (Greenawalt et al. \markcite{gwb97}1997).  Ferguson et
al. (\markcite{fwg96b}1996a) have also seen a similar continuous range
of $\rm I_{[SII]}/I_{H\alpha}$ ratios from DIG to HII regions in the
edge-on galaxy NGC 55.  These observed trends would suggest that DIG
and HII regions are intimately related by photoionization, and are
actually two extremes in a continuous class of warm ionized gas
structures.  However, it is not clear why a similar trend of
decreasing $\rm I_{[OIII]}/I_{H\alpha}$ with increasing surface
brightness is not seen as predicted by the DM models.

\subsection{Mixing Layers}

The turbulent mixing of hot and cool gas to produce warm gas at the
boundary has been used by Slavin et al. \markcite{ssb93}(1993)
(hereafter, SSB) to model some of the DIG in the Galaxy.  They
proposed that energy input from supernovae could be responsible for
ionizing roughly 20\% of the Galactic DIG.  Hundreds of mixing layers
along any given line of sight are thought to account for the observed
surface brightness of ionized gas.  The mixing speed of the hot gas
and the temperature attained by the warm gas are important in the
model.  We compare to their Solar abundance models because the
depleted abundance models predict elevated [SII] line strengths which
do not fit the observations as well.  The predictions for models with
log($\rm\bar{T}$)= 5.0, 5.3, and 5.5 are denoted with an ``S'' in figs
4 \& 5.  At each temperature, two mixing speeds are presented (v = 25
\& 100 $\rm km~s^{-1}$).  Attributing all emission to mixing layers is
not plausible, because an extremely specific choice of parameters is
required to match the observations.  As can be seen in figs 4 \& 5,
for the log($\rm\bar{T}$)= 5.3 \& 5.5 models, the [OIII] predictions
are an order of magnitude above the observed values, while the [OII]
line is a factor of 5 above the observed values.  Lowering the
temperature decreases the oxygen line strengths, but a small change in
temperature produces a very drastic decrease in the line strengths.
Lowering the temperature to log($\rm\bar{T}$)= 5.0 (a factor of 2 or
3), brings the [OII] strength to a comparable level to the
observations.  However, the [OIII] line strength drops by two orders
of magnitude, so that it is now seriously underpredicted, as is the
[SII] line strength.  Given the extreme sensitivity of model
predictions to this temperature range, a very specific set of model
parameters would be required to fit the observations.  However, such a
specific choice of parameters may be unlikely to be valid for many DIG
regions spread over a large portion of a spiral galaxy, which is what
comprises our representative DIG spectra.  Therefore, regardless of
the energetic arguments, this model in itself appears unlikely to
account for all observed DIG.

Given the above arguments, turbulent mixing is likely only responsible
for a small fraction of the DIG ionization, with photoionization being
the dominant source.  Therefore, a mixture of the SSB and DM models
may be a more accurate way of modelling the observed line ratios.
Given the predicted line ratios for the DM and SSB models it is
straightforward to determine the predicted line ratios for a model
which attributes a certain fraction of the ionization of the DIG to
turbulent mixing layers and the remaining amount to photoionization.
Specifically, the predicted line ratio will be $R=\Sigma f_{i}R_{i}$,
where $R_{i}$ is the predicted line ratio for a component that
contributions the fraction $f_{i}$ to the total ionization.  We
calculated line ratios for models that combined either the SSB
log($\rm\bar{T}$)= 5.0, v=100 km s$^{-1}$ model (labeled S$_{2}$ in
figs. 4 \& 5) or the log($\rm\bar{T}$)= 5.3, v=25 km s$^{-1}$ model
(labeled S$_{3}$) with the DM X$_{edge}$=.1 model with q=.005 (tick
mark 6 in figs. 4 \& 5).  Calculations were made for 5\%, 10\%, 20\%,
and 30\% contribution from turbulent mixing.  Turbulent mixing at v=25
km s$^{-1}$ that results in gas with log($\rm\bar{T}$)= 5.3 is
unlikely to contribute more than 5\% to the ``Very Faint'' DIG in M31
because the predicted $\rm I_{[OIII]}/I_{H\alpha}$ ratio would be well
above our observations.  The predictions for this model do not match
any other observations, except possibly the ``Moderate'' DIG if the
contribution from turbulent mixing is at the few percent level.
Turbulent mixing at v=100 km s$^{-1}$, resulting in gas with
log($\rm\bar{T}$)= 5.0, can contribute up to 20\% and still be
consistent with our observation of the DIG in M31.  As can be seen in
fig. 4, although up to 20\% of the ionization can arise from turbulent
mixing, smaller contributions fit our observations better.  Models that
attributed the remaining fraction of ionization to DM's ``composite''
model with log(q)=-3.7 instead of the X$_{edge}$=.1 model predicted
$\rm I_{[SII]}/I_{H\alpha}$, $\rm I_{[OIII]}/I_{H\alpha}$, and $\rm
I_{[OII]}/I_{H\alpha}$ ratios that did not fit any observations.

\subsection{Discussion}

Recent spectroscopy of NGC 891 (Rand 1996) has shown that $\rm
I_{[NII]}/I_{H\alpha}$ varies from a value of .4, in the plane, to
about 1.4, 3.5 kpc out of the plane.  This is quite different than
similar observations for NGC 4631 (Golla et al. 1996) which find that
the ratio is never greater than .6 and decreases to .2 in the
midplane.  Comparison of our observations to this extraplanar ionized
gas suggests that the bulk of the emission from the DIG we see in M31
is at or near the midplane.  In M31, the $\rm I_{[NII]}/I_{H\alpha}$
ratio for all DIG brightness levels is .3-.35.  As this is near the
values obtained for the midplane gas in edge-on galaxies, it is
reasonable to believe that the bulk of the DIG in M31 is likely
relatively close to the plane.  Similarly, the ratios measured for
$\rm I_{[OII]}/I_{H\alpha}$ in M31 are near the low end of the range
of values determined by recent narrow band [OII] imaging of NGC 55
(Ferguson et al. 1996a), roughly .7 in the midplane to as high as 2 at
about 1kpc above the disk.  This also supports the idea that the DIG
observed in M31 has spectral properties consistent with DIG near the
midplane of edge-on galaxies, although differences in abundance
between M31 and NGC 55 may play a role as well.

Several emission lines have been observed for the Reynolds layer of
the Milky Way (eg. Reynolds
\markcite{rey84}1984,\markcite{rey85a}1985a,\markcite{rey85b}1985b,
Reynolds \& Tufte \markcite{rt95}1995).  They are included in Table 1.
The emission measures of most of these observations are in the same
range as the observations we present here for M31.  Likewise, some of
the observed line ratios occupy the same range of values.  For
instance, $\rm I_{[SII]}/I_{H\alpha}$ is generally observed to be
higher in the Milky Way, but it is sometimes seen as low, or lower, as
in M31.  Noted exceptions are the ratios $\rm I_{[OIII]}/I_{H\beta}$
and $\rm I_{HeI}/I_{H\alpha}$, which appear lower in the Galactic DIG
than in M31.  Photoionization models account for these differences in
line ratios by requiring the DIG in the Galaxy to be subjected to a
more diluted radiation field in comparison to the the DIG in M31.
Given that the DIG we observed in M31 is located in or near the spiral
arms, it is reasonable to assume that the stellar radiation field may
be harder in comparison to Galactic DIG regions due to a higher
fraction of direct stellar versus diffuse radiation.  In addition,
higher q could imply either higher intensity or somewhat lower $n_e$;
lower $n_e$ is not likely (see Walterbos \& Braun
\markcite{wb94}1994).  Since the emission measures are similar for
both sets of DIG, Galactic and M31, it may be that different
photoionization environments are important in different DIG regions
within a galaxy.  Testing this hypothesis would require the analysis
of DIG regions between and within the spiral arms of a nearby spiral
galaxy looking for differences in spectroscopic properties at the same
DIG brightness level.

\section{Conclusions}

We have obtained sensitive spectra of representative DIG within M31 at
four surface brightness levels, ranging from 9-59 pc cm$^{-6}$, and of
DIG specifically selected ``far'' from HII regions.  From these
spectra, we have measured H$\alpha$, H$\beta$, H$\gamma$, H$\delta$,
[NII]$\lambda6583$, [SII]($\lambda6717+\lambda 6731$),
[OIII]$\lambda5007$, and [OII]$\lambda3727$ line intensities.  Our
data represent a more complete set of line observations for DIG
within the disk of an external galaxy than has been available before.

Our results are:

\begin{itemize}

\item{We tentatively detect HeI$\lambda5876$ in the average DIG
spectra with EM = 58.4 pc cm$^{-6}$.  The observed ratio is $\rm
I_{HeI}/I_{H\alpha} = .045\pm.015$ implying that helium is nearly
100\% singly ionized.  This is in contrast to the upper limits
detected for Galactic DIG, which suggest that helium is predominantly
neutral.  The M31 DIG regions are located primarily near prominent
spiral arms, while the Galactic DIG regions are not.  One would expect
that within DIG regions immediately surrounding a spiral arm, the
stellar radiation field should be harder compared to DIG regions away
from spiral arms, in agreement with our results.  For fainter DIG, our
He line strength upper limits are consistent with the He being either
ionized or neutral. }

\item{The photoionization models of Domg$\rm\ddot o$rgen \& Mathis
\markcite{dm94}(1994), appropriate for DIG within the Galaxy,
underpredict the [OIII] flux while overpredicting the [SII] flux for
DIG in M31.  However, a higher q model, where q characterizes the
relative importance of the stellar radiation field compared to the
diffuse nebular field, is able to reproduce the observations.  The
need for the larger q value to fit our observations may not be
suprising given that the M31 DIG regions are primarily located near
spiral arms. }

\item{The observed forbidden line ratios suggest that turbulent mixing
(Slavin et al. \markcite{ssb93}1993) appears to contribute only a
small fraction of the ionization of the DIG.  In particular, models
that attribute a fraction of the ionization to log($\rm\bar{T}$)= 5.3,
v=25 km s$^{-1}$ turbulent mixing and the remaining amount to
photoionization, can reasonably well fit the observations of the
``Very Faint'' DIG if the contribution from turbulent mixing is at
most about 5\%. The predicted line ratios do not match line ratios for
brighter DIG levels discussed here.  Conversely, log($\rm\bar{T}$)=
5.0, v=100 km s$^{-1}$ turbulent mixing can contribute at most 20\% of
the ionization of DIG, but lower fractions give better fits.}

\end{itemize}

We thank J. Mathis for discussion regarding the subject of this paper.
This work has been supported by a grant from NSF (AST 9123777) to
R. A. M. W., and a grant to B. G. from the New Mexico Space Grant
Consortium.  We also acknowledge support from the Research Corporation
through a Cottrell Scholarship grant to R. A. M. W.

\clearpage

\clearpage
\begin{figure}
\plotfiddle{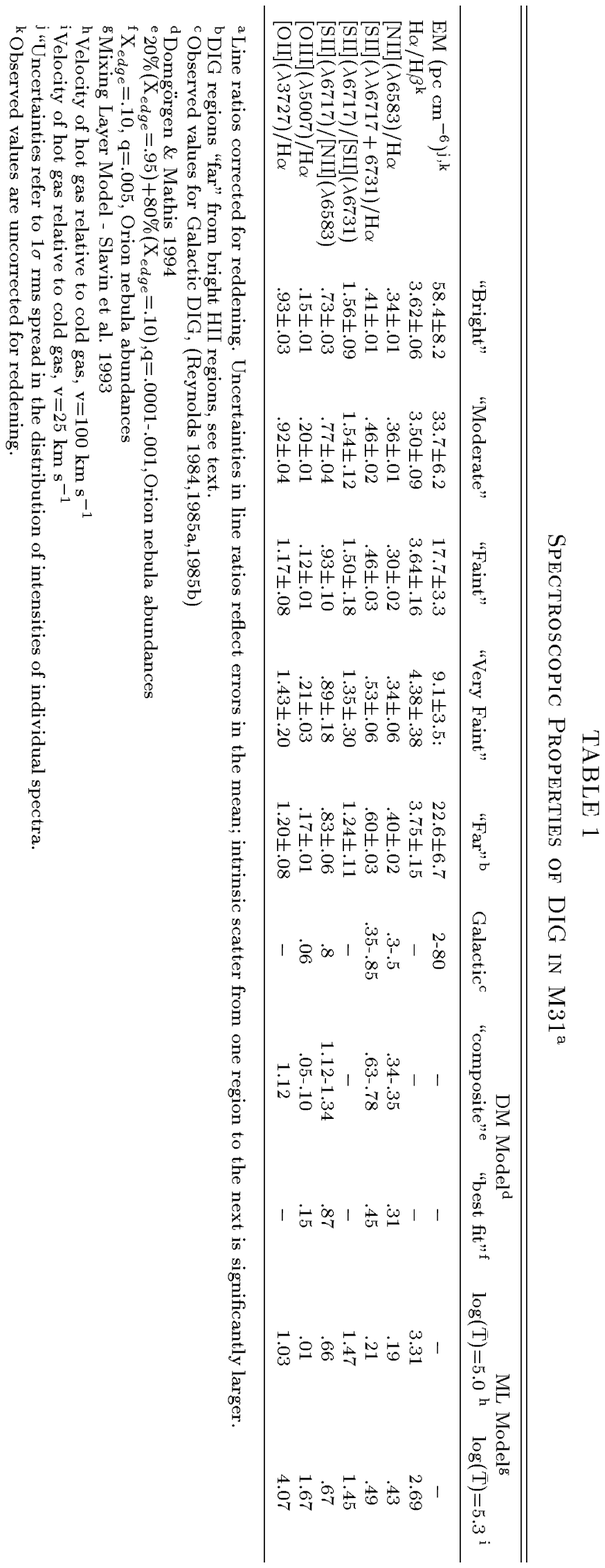}{8in}{0}{95}{90}{-200}{-90}
\end{figure}
\clearpage

\begin{figure}
\plotfiddle{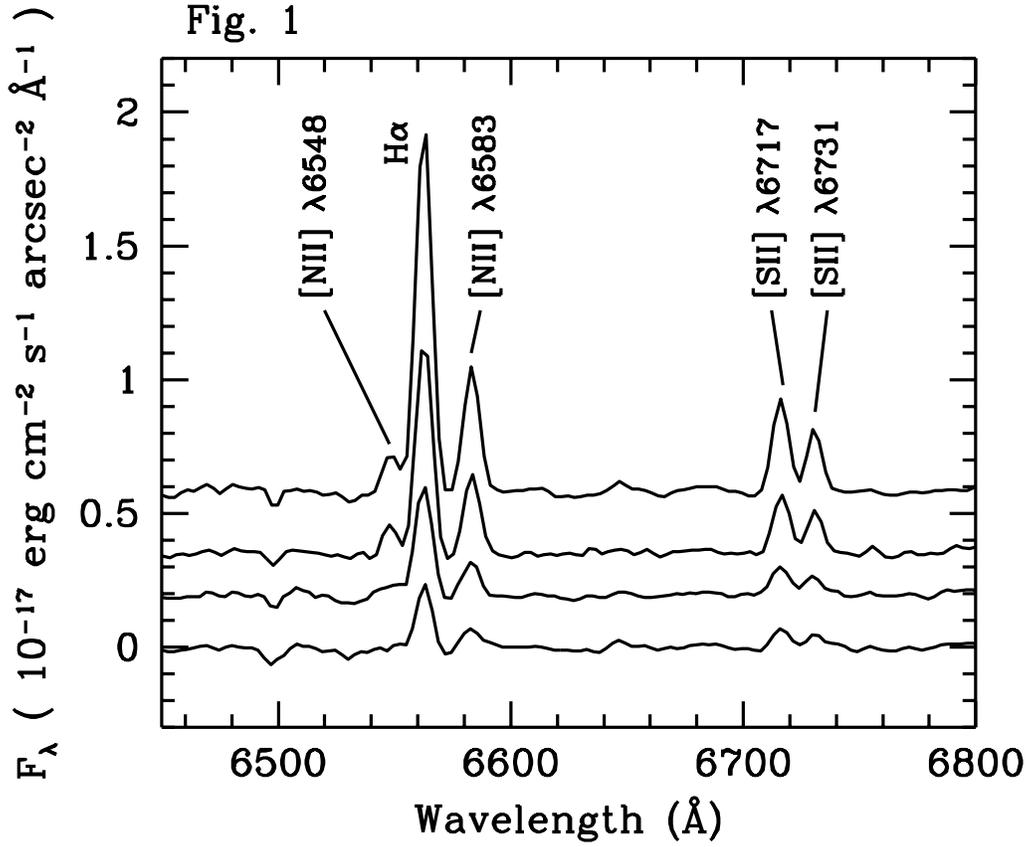}{6in}{0}{120}{120}{-360}{-300}
\figcaption[fig1.eps]{ Co-added red spectra of representative DIG
regions in M31.  This spectral region includes lines of H$\alpha$,
[NII]$(\lambda 6548+\lambda 6583)$, and [SII]$(\lambda 6717+\lambda
6731)$.  The spectra shown, from top to bottom, are the ``Bright'',
``Moderate'', ``Faint'' and ``Very Faint'' DIG.  A constant offset was
added to the ``Bright'', ``Moderate'', and ``Faint'' spectra.  Many
night sky lines blueward of H$\alpha$ cause problems with the
[NII]$\lambda 6548$ line determination and the dip in the baseline at
$\rm \sim6500\AA$.  The [SII] emission is enhanced in DIG compared to
HII regions, while [NII] is at roughly the same level.  All DIG
spectra were shifted to account for the systemic and rotational
velocity of M31 before co-adding.}
\end{figure}

\begin{figure}
\plotfiddle{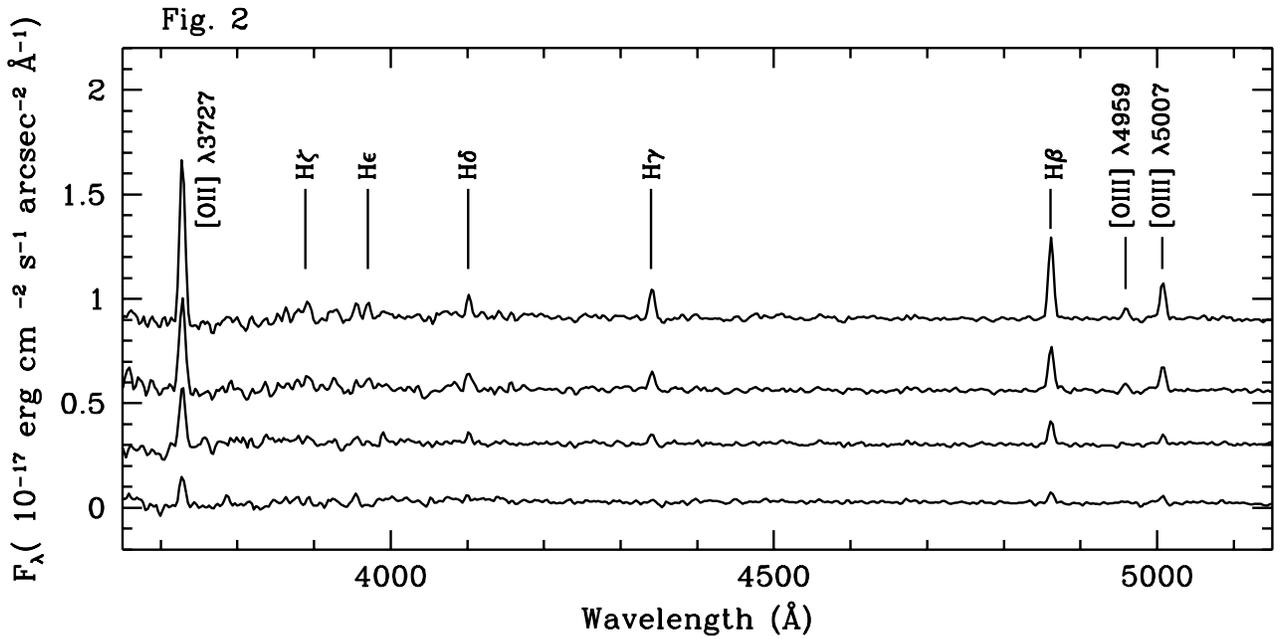}{6in}{0}{90}{90}{-270}{-200}
\figcaption[fig2.eps]{ Co-added blue spectra of representative DIG
regions in M31.  Only the ``blue'' spectra contain observations
blueward of 4000$\rm\AA$, thus causing the decreased signal-to-noise
(see text).  The spectra shown, from top to bottom, are the
``Bright'', ``Moderate'', ``Faint'', and ``Very Faint'' DIG.  A
constant offset was added to the ``Bright'', ``Moderate'', and
``Faint'' spectra.  Besides the hydrogen lines in this region, we have
clear detections of [OII]$\lambda 3727$, and [OIII]$\lambda 5007$ in
all 4 DIG brightness levels.  All DIG spectra have been shifted to
account for the systemic and rotational velocity of M31 before
co-adding.}
\end{figure}

\begin{figure}
\plotfiddle{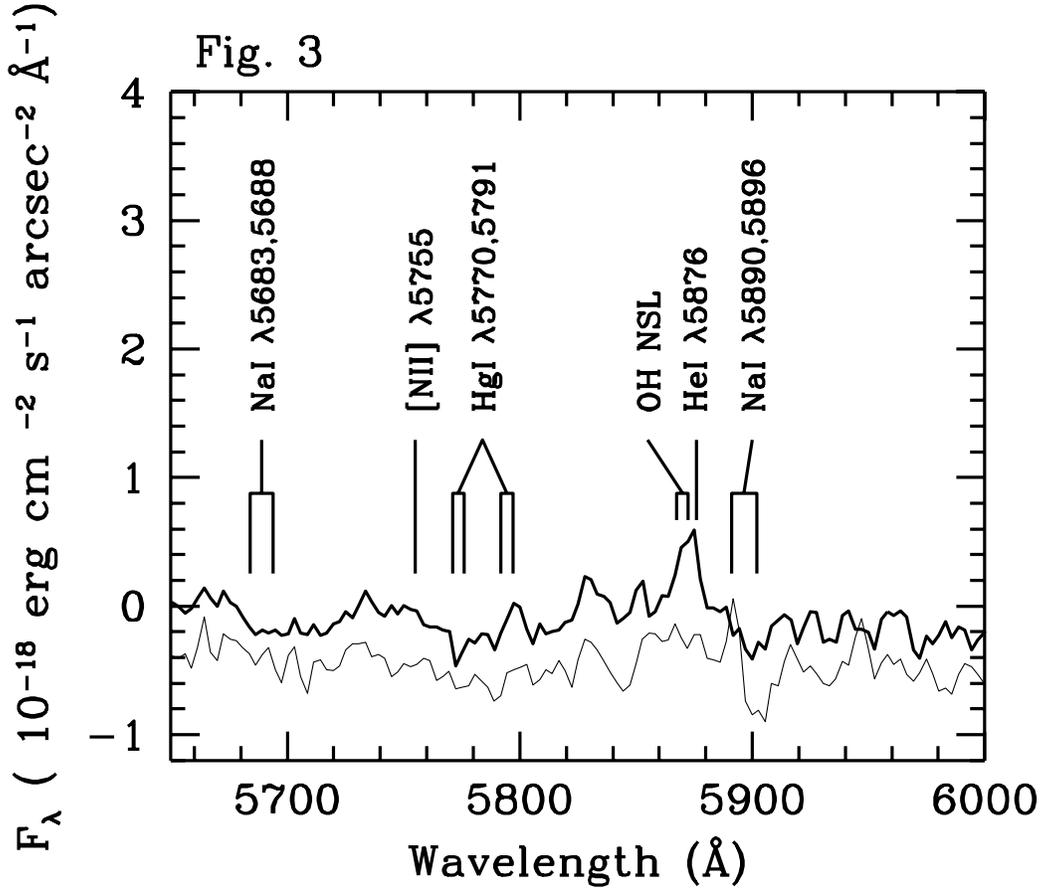}{6in}{0}{120}{120}{-360}{-300}
\figcaption[fig3.eps]{ Observed ``red'' spectra of representative DIG
in M31.  The thick lined spectrum is the ``Bright'' DIG, while the
thin lined spectrum is the ``Moderate'' DIG.  The locations of the HeI
and the [NII]$\lambda 5755$ lines are shown along with the location of
several important night sky lines in this spectral region.  There is
some correlated ``noise'' between the two spectra because of night
sky line residuals.}
\end{figure}

\begin{figure}
\plotone{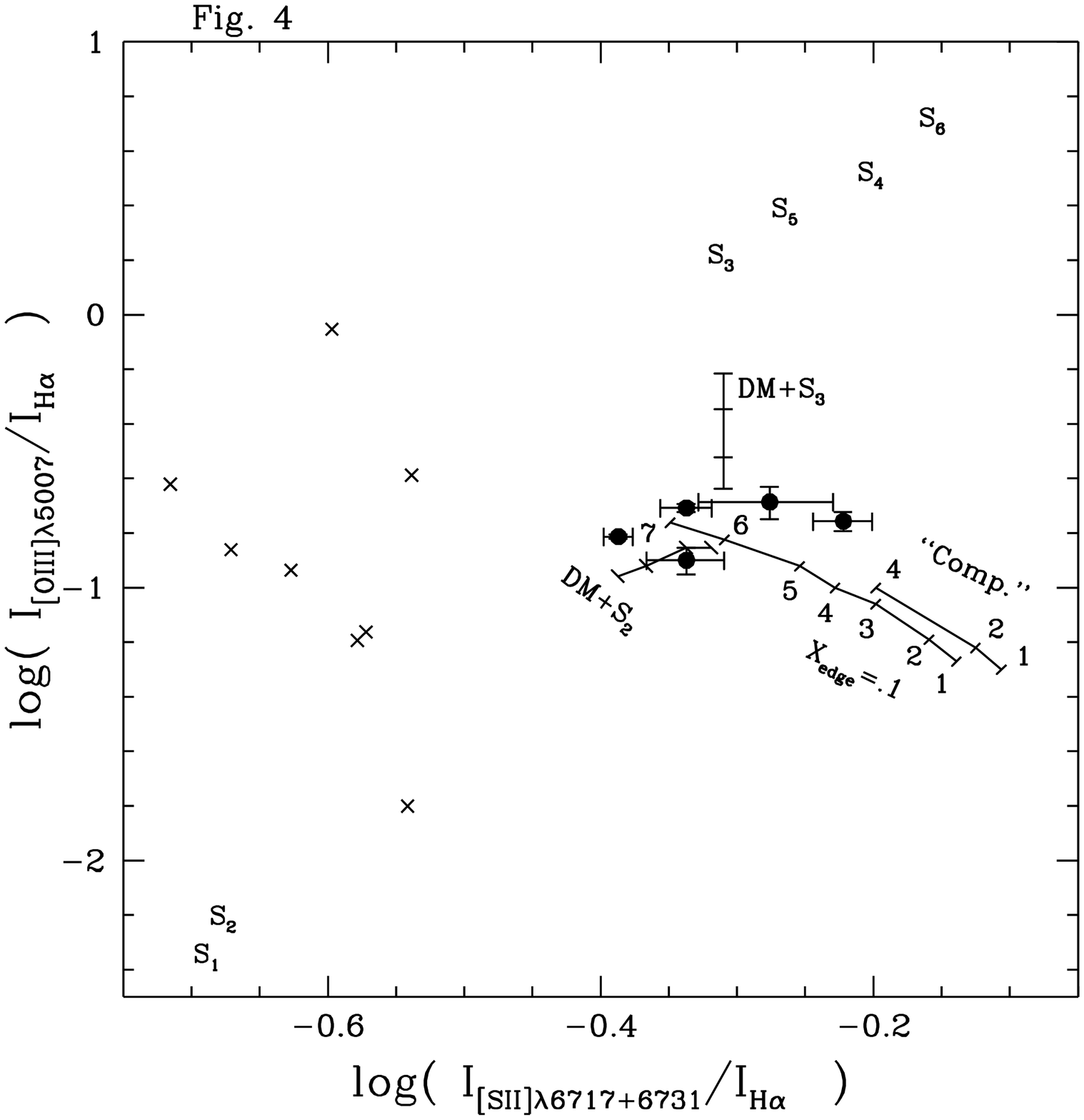}
\figcaption[fig4.eps]{ $I_{[SII]}/I_{H\alpha}$
vs. $I_{[OIII]}/I_{H\alpha}$ for representative DIG (filled circles) in
M31.  The $\times$'s are some HII regions from M31 (Galarza et
al. 1997).  The loci of line ratio predictions from the Domg$\rm\ddot
o$rgen \& Mathis (1994) models are label ``Comp.'' and X$_{edge}$=.1;
the tick mark labels 1-7 refer to the log(q) values of
-4,-3.7,-3.3,-3,-2.7,-2.3,-2, respectively.  The mixing layer models
with solar abundance (Slavin et al. 1993) are labeled with S's.  The
subscripts 1,2 correspond to log($\rm\bar{T}$)= 5.0 models, while 3,4
and 5,6 correspond to log($\rm\bar{T}$)= 5.3 \& 5.5 models
respectively.  Odd subscripts are for models of mixing speed 25 km
s$^{-1}$, while even subscripts are for 100 km s$^{-1}$ models.
Finally, the loci of predictions for the combined
photoionization/mixing layer models discussed in the text are labeled
DM+S$_{2}$ and DM+S$_{3}$.  Tick marks correspond to 5\%, 10\%, 20\%,
and 30\% contribution from mixing layers.}
\end{figure}

\begin{figure}
\plotone{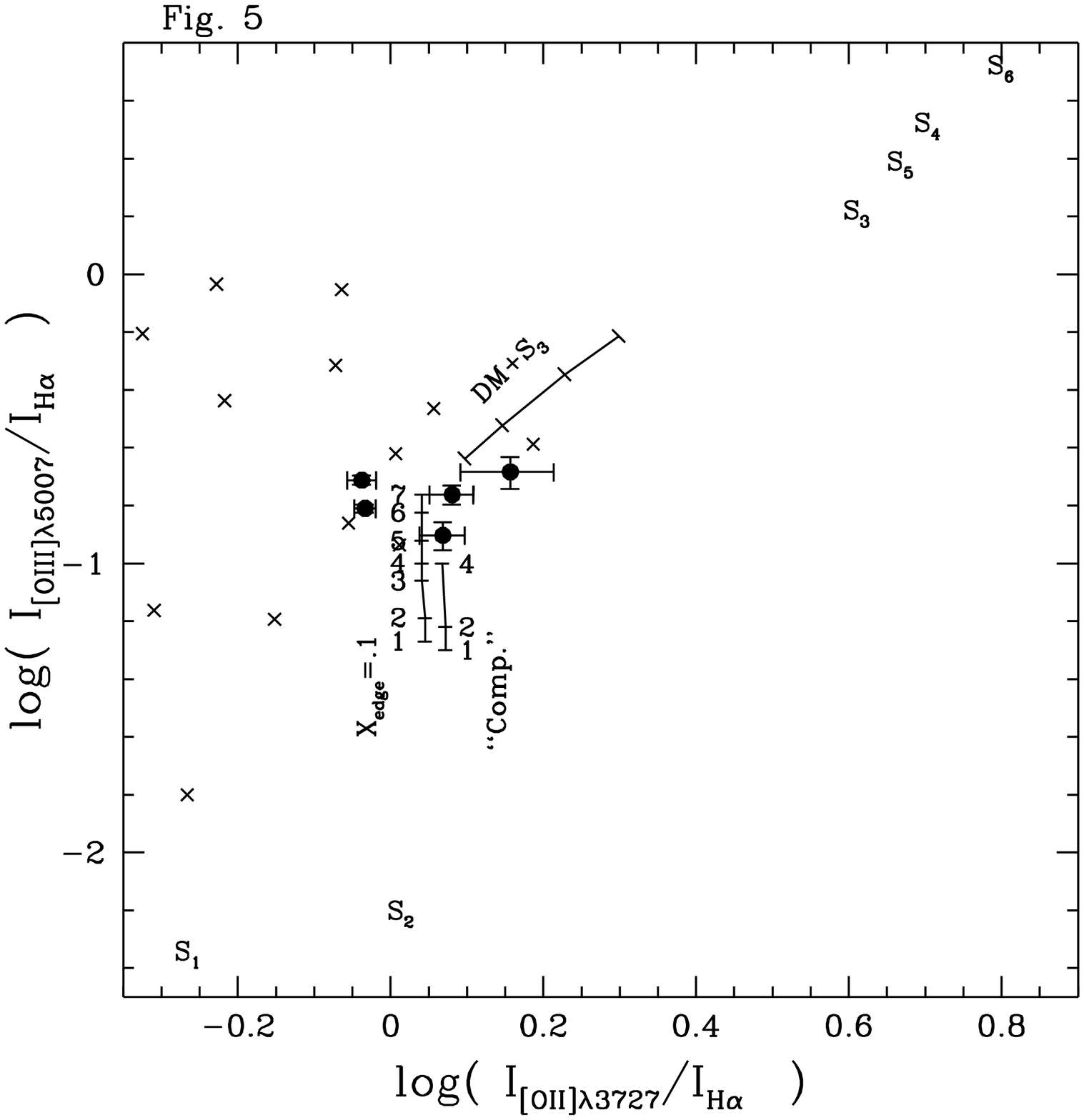}
\figcaption[fig5.eps]{ $I_{[OII]}/I_{H\alpha}$
vs. $I_{[OIII]}/I_{H\alpha}$ for representative DIG (filled circles)
in M31.  The $\times$'s are some HII regions in M31 (Galarza et
al. 1997).  The loci of line ratio predictions from the Domg$\rm\ddot
o$rgen \& Mathis (1994) models are label ``Comp.'' and X$_{edge}$=.1;
the tick mark labels 1-7 refer to the log(q) values of
-4,-3.7,-3.3,-3,-2.7,-2.3,-2, respectively.  The mixing layer models
with solar abundance (Slavin et al. 1993) are labeled with S's.  The
subscripts 1,2 correspond to log($\rm\bar{T}$)= 5.0 models, while 3,4
and 5,6 correspond to log($\rm\bar{T}$)= 5.3 \& 5.5 models
respectively.  Odd subscripts are for models of mixing speed 25 km
s$^{-1}$, while even subscripts are for 100 km s$^{-1}$ models.
Finally, the locus of predictions for the combined
photoionization/mixing layer models discussed in the text is labeled
DM+S$_{3}$.  Tick marks correspond to 5\%, 10\%, and 20\% contribution
from mixing layers.  The results for DM+S$_2$ are not shown because
they overlap the DM X$_{edge}$=.1 model predictions and fall between
tick marks 4-6. }
\end{figure}

\clearpage


\begin{references}

\reference{det90} Dettmar, R.-J. 1990, \aap, 232, L15
\reference{det92} Dettmar, R.-J. 1992, Fund. Cos. Phys., 15, 143
\reference{ds92} Dettmar, R.-J. \& Schulz, H. 1992, \apjl, 254,  L25
\reference{dm94} Domg$\rm\ddot o$rgen, H. and Mathis, J.S. 1994, \apj,
428, 647 
\reference{ds94} Dove, J.B. \& Shull, J.M., 1994, \apj, 430, 222
\reference{fwg96a} Ferguson, A.M.N., Wyse, R.F.G., and Gallagher, J.S. 1996a, astro-ph/9609125 
\reference{fwgh96b} Ferguson, A.M.N., Wyse, R.F.G., Gallagher, J.S., and Hunter, D.A. 1996b, \aj, 111, 2265
\reference{gwg97} Galarza, V., Walterbos, R.A.M., Greenawalt, B.E., and Braun, R. 1997, in preparation 
\reference{gdd96} Golla, G., Dettmar, R.-J., and Domg$\rm\ddot o$rgen, H. 1996, \aap, 313, 439
\reference{gw97a} Greenawalt, B.E., \& Walterbos, R.A.M. 1997a, in preparation
\reference{gw97b} Greenawalt, B.E., \& Walterbos, R.A.M. 1997b, in preparation
\reference{gwb97} Greenawalt, B.E., Walterbos, R.A.M., and Braun, R. 1997, in
preparation 
\reference{hk90} Hester, J.J., \& Kulkarni, S.R. 1990, in The Interstellar Medium in External Galaxies, edited by D. Hollenbach and H. Thronsen (NASA CP-3084), p.288
\reference{hwg96} Hoopes, C.G., Walterbos, R.A.M., and Greenawalt, B.E. 1996, \aj, 112, 1429
\reference{hunt94a} Hunter, D.A. 1994a, \aj, 107, 565
\reference{hunt94b} Hunter, D.A. 1994b, \aj, 108, 1658
\reference{hunt96} Hunter, D.A. 1996, \apj, 457, 671
\reference{hg90} Hunter, D.A., \& Gallagher, J.S. 1990, \apj, 362, 480
\reference{hg92} Hunter, D.A., \& Gallagher, J.S. 1992, \apjl, 391, L9
\reference{hg96} Hunter, D.A., \& Gallagher, J.S. 1996, preprint 
\reference{hhg93} Hunter, D.A., Hawley, W.N., \& Gallagher, J.S. 1993, \aj, 106, 1797
\reference{kdgr91} Keppel, J.W., Dettmar, R.-J., Gallagher, J.S., \& Roberts, M.S. 1991, \apj, 374, 507
\reference{kh88} Kulkarni, S.R., \& Heiles, C.E. 1988, in Galactic and Extragalactic Radio Astronomy, edited by G.L. Verschuur and K.I. Kellerman (Springer, New York), p. 95
\reference{math86} Mathis, J.S. 1986, \apj, 301, 423
\reference{math96} Mathis, J.S. 1996, private communication
\reference{mc93} Miller, W.W. \& Cox, D.P. 1993, \apj, 417, 579
\reference{ost89} Osterbrock, D.E. 1989, Astrophysics of Gaseous Nebulae and Active Galactic Nuclei (Mill Valley: University Science Books)
\reference{rand94} Rand, R.J. 1994, in Proceedings of IAU Joint Discussion 1, ed. van der Hulst, J.M. 
\reference{rand96} Rand, R.J. 1996, preprint 
\reference{rkh90} Rand, R.J., Kulkarni, S.R., \& Hester, J.J. 1990, \apjl, 352, L1
\reference{rkh92} Rand, R.J., Kulkarni, S.R., \& Hester, J.J. 1992, \apj, 396, 97
\reference{ray79} Raymond, J.C. 1979, \apjs, 39, 1
\reference{rey84} Reynolds, R.J. 1984, \apj, 282, 191
\reference{rey85a} Reynolds, R.J. 1985a, \apj, 294, 256
\reference{rey85b} Reynolds, R.J. 1985b, \apj, 298, L27 
\reference{rey90} Reynolds, R.J. 1990, in The Galactic and Extragalactic Background Radiation, IAU Symposium No. 139, edited by S. Bowyer and C. Leinert (Kluwer, Dordrecht), p. 157
\reference{rey91} Reynolds, R.J. 1991, \apjl, 372, L17
\reference{rt95} Reynolds, R.J. \& Tufte, S.L. 1995, \apjl, 439, L17
\reference{sm79} Savage, B.D. \& Mathis, J.S. 1979, ARA\&A, 17, 73
\reference{sci90} Sciama, D.W. 1990, \apj, 364, 549
\reference{sci93} Sciama, D.W. 1993, \apjl, 409, L25
\reference{sci95} Sciama, D.W. 1995, \apj, 448, 667
\reference{s79} Shull, J.M. \& McKee, C. 1979, \apj, 227, 131
\reference{ssb93} Slavin, J.D., Shull, J.M., and Begelman, M.C. 1993, \apj, 407, 83
\reference{tlp74} Torres-Peimbert, S., Lazcano-Araujo, A., \& Peimbert, M. 1974, \apj, 191, 401
\reference{wb92} Walterbos, R.A.M. \& Braun, R. 1992, \aaps, 92, 625 
\reference{wb94} Walterbos, R.A.M. \& Braun, R. 1994, \apj, 431, 156
\reference{wkb96} Walterbos, R.A.M., King, N., \& Braun, R. 1996, submitted to \apjl
\reference{ws87} Walterbos, R.A.M. \& Schwering, P.B.W. 1987, \aaps, 66, 505

\end{references}
\end{document}